\documentclass[a4paper]{article}

\usepackage[english]{babel}
\usepackage[utf8]{inputenc}
\usepackage{braket} 
\usepackage{physics} 
\usepackage{amsmath}
\usepackage{graphicx}
\usepackage[colorinlistoftodos]{todonotes}

\title{Ordering the processes with indefinite causal order}

\author{Stanislav Filatov and Marcis Auzinsh\\
\textit{University of Latvia, Department of Physics}\\
\textit{Raina boulevard 19, LV-1586, Riga, Latvia}\\
              E-mail: sutasu@tutanota.com
             }

\date{\today}

\begin{document}
\maketitle

\begin{abstract}

We show a method of describing processes with indefinite causal order (ICO) by a definite causal order. We do so by relabeling the processes that take place in the circuit in accordance with the basis of measurement of control qubit. Causal nonseparability is aleviated at a cost of nonlocality of the acting processes. This result highlights the key role of superposition in creating the paradox of ICO. We also draw attention to the issue of growing incompatibility of language in its current form (especially the logical structures it embodies) with the quantum logic. 

\end{abstract}

\maketitle 

\section{Introduction}

Indefinite Causal Order (ICO) implemented through Quantum Switch has seen a lot of attention recently \cite{1,2,3}. Apart from practical advances \cite{4,5,6} the approach implicitly claims a fundamental point as well: ICO setups, being inherently unorderable, hint at the inherent unorderability and acausality of quantum mechanics and therefore the nature. We would like to refine this point and show that it is not valid universally, but only within a particularly chosen reference frame. 

First of all we consider the case of a particle being in spatial superposition and show in what sense it occupies two places simultaneously and in what sense it is in a particular place. Then we expand the analogy to the case of ICO processes and show how to order them. Finally, remarks about linguistic character of the difficulties of locating and ordering the superpositions in quantum mechanics are presented along with conclusions. We use ordinary-language descriptions along with technical throughout the text in order to highlight the ways in which the paradox of acausality arises.

\section{Analogy with spatial superposition}

Let us look at a particle entering an interferometer which has two arms. Depending on the conditions (state of the particle, setup of an interferometer) particle may take the arm \textbf{A}, arm \textbf{B} or a superposition of arms (we use \textbf{bold} to depict the names of the arms)\footnote{We will not be using bold in some descriptions when the emphasis of naming is not that important}. When conditions are right for the particle to take the arm \textbf{A}, we say ``Particle is in arm \textbf{A} and not in arm \textbf{B}". In language of quantum mechanics we may say ``The state of the system is $\ket{A}$ and not the one orthogonal to it, $\ket{B}$". When the conditions are right for the particle to take a superpositional route, say $\ket{(A+B)/\sqrt{2}}$, we may be tempted to say that particle takes two arms simultaneously and that there is no particular ``arm" where particle is. Nevertheless, in the language of quantum mechanics there is precise expression for the state of the system ``The state of the system is $\ket{(A+B)/\sqrt{2}}$ and not the one orthogonal to it, $\ket{(A-B)/\sqrt{2}}$". Translating this to our language we may say ``Particle is in arm $\mathbf{(A+B)/\sqrt{2}}$ and not in arm $\mathbf{(A-B)/\sqrt{2}}$."\footnote{This sentence should be read as ``Particle is in arm ``A plus B divided by square root 2" and not in arm ``A minus B divided by square root 2."" Text in bold denotes the name of the arm.} 

We would like to stress that just like in the first example, in the second example there is particular ``place" or ``arm" in which the particle definitely is and particular ``place" or ``arm" in which the particle definitely is not. That existence may be confirmed experimentally in both cases. The difference between the examples is that our everyday macroscopic intuition allows to call localized arm \textbf{A} an ``arm", but does not allow to call delocalized arm $\mathbf{(A+B)/\sqrt{2}}$ an ``arm". In the next section we will show how this analogy may be applied to ICO processes. 

\section{Ordering the ICO}

We assume familiarity of reader with general properties of ICO and Quantum Switch, for a detailed description of ICO processes one is referred to the aforementioned literature, here we will concentrate on the key moments. Instead of two arms \textbf{A} and \textbf{B}, there are two processes \textbf{A} and \textbf{B} through which a particle might pass. Moreover, the processes are connected between each other so that A may happen first, then B, then final measurement of the attributes of the particle; or B first, then A, then final measurement. Important, although A and B are arbitrary processes (that might even include measurement and collapse in a sense), they are connected coherently, i.e. A and B don't involve irreversible processes (see the descriptions of experimental setups in the cited literature). Finally, the order ``A then B" vs ``B then A" is related to the state of control qubit (usually degree of freedom of the same particle that goes through the circuit). When the state of control qubit is in superposition between the states corresponding to particular order, indefinite causal order of processes A and B appears because the particle now simultaneously passes through ``A then B" and ``B then A" processes.

For clarity of our argument we restrict the ``processes" we are referring to in this work to unitary operations. Although extension of our argument onto Completely Positive Trace-preserving maps using Stinespring dilation \cite{65} seems possible, it would require even more non-intuitive relabelling of the processes and would clout the essence of our reasoning. We leave this extension for the future.

In order to untangle the situation one should note that, although the processes A and B might be very complex, essentially there are just two possibilities of what might happen ``A then B" or ``B then A". Moreover, two possibilities are absolutely determined by what process happens first. Namely, ``A then B" could be described as ``A happens first". Because the system consists of just two processes, and the particle passes through both, this statement is enough to deduce the whole situation ``A then B". 

Before we start employing BraKet-notation for processes we should describe the Hilbert-space in which they reside. We follow the authors of original ICO scheme \cite{2} and researchers of superpositions of quantum evolutions \cite{7} in applying the Choi-Jamiolkowski isomorphism \cite{8,9} to obtain that Hilbert space. The isomorphism represents the linear operator mapping states in input Hilbert-space $\mathcal{H}_{in}$ onto states in output Hilbert-space $\mathcal{H}_{out}$ as a vector in Hilbert-space $\mathcal{H}_{out}\otimes\mathcal{H}_{out}$. Then $\ket{A}$ (``process named A") will reside in H-space (``pr" stands for ``process") 

\begin{align}
\mathcal{H}_{pr}^{A}= \mathcal{H}_{in}^{A}\otimes\mathcal{H}_{out}^{A}
\end{align}

$\ket{B}$ (``process named B") will reside in H-space

\begin{align}
\mathcal{H}_{pr}^{B}= \mathcal{H}_{in}^{B}\otimes\mathcal{H}_{out}^{B}
\end{align}

And therefore both processes $\ket{A}$ and $\ket{B}$ will reside in H-space 

\begin{align}
\mathcal{H}_{pr}^{AB}= \mathcal{H}_{pr}^{A}\otimes\mathcal{H}_{pr}^{B}
\end{align}

We could, if we wanted to, shrink the H-space to the one spanned by $\ket{A}$ and $\ket{B}$ to obtain 2-dimensional H-space reflecting two possible answers to the question ``Which process hapenned first?" 

We should also mention orthogonality of $\ket{A}$ and $\ket{B}$ that is assumed in this analysis. To order the processes those processes must be different (in language of quantum mechanics the vectors representing them must be orthogonal). That is why there is no sense of order to the same process repeated twice and hence no need of ordering. This line of reasoning doesn't rule out the case of partial overlap. We must admit that we don't have a rigorous proof why such cases may or may not always be reduced to a combination of aforementioned extreme cases of zero and full overlap. But we have a feeling that it can be done in the following way. 

Let us say $\ket{V}$ and $\ket{W}$ are mutually non-orthogonal representations of some processes V and W. Let's say V precedes W. Such a description will not be an accurate ordering of the processes because there is a part in each of V and W that doesn't have a sense of orderedness to it. This part is projection Proj($\ket{V};\ket{W}$). Subtracting the projection from one of the vectors and scaling them appropriately we obtain two orthogonal vectors that have notion of orderedness applicable to them. The apparent order of the processes in our circuit is only applicable to these distilled and mutually orthogonal vectors.

Now let us try to describe the ordinary-language situation ``first \textbf{A} then \textbf{B}" situation in quantum language ``$\ket{A}$ happens first, not its orthogonal $\ket{B}$, then $\ket{B}$ happens, not its orthogonal $\ket{A}$". Situation ``first \textbf{B} then \textbf{A}" is described analogously. In order to describe in a similar way the superpositional case we need to look at the control qubit, because its state fully determines the path of the particle through the processes A and B. Let us say that state $\ket{a}$ of control qubit corresponds to ``first \textbf{A} then \textbf{B}", other way around for $\ket{b}$. In other words, the state of control qubit uniquely determines the first process through which the particle undergoes. Then the state of control qubit $\ket{(a+b)/\sqrt{2}}$ also uniquely determines the first process through which the particle undergoes. Let us say it is $\ket{(A+B)/\sqrt{2}}$. The first process (as we have discussed) uniquely determines the second process - which is orthogonal to it (in this case $\ket{(A-B)/\sqrt{2}}$). In quantum language we would say ``$\ket{(A+B)/\sqrt{2}}$ happens first, not its orthogonal $\ket{(A-B)/\sqrt{2}}$, then $\ket{(A-B)/\sqrt{2}}$ happens, not its orthogonal $\ket{(A+B)/\sqrt{2}}$". That translates to ordinary language ``first $\mathbf{(A+B)/\sqrt{2}}$ then $\mathbf{(A-B)/\sqrt{2}}$". This is essentially the recipe for ordering ICO processes: following the whole particle through the circuit and noticing that superpositional states of control qubits uniquely determine paths in Hilbert-space. Extensions to the circuits with more than two processes should be possible as well. 

Interpretation of the meaning of the new name of the proces being $\mathbf{(A+B)/\sqrt{2}}$ beyond realization that it is a superposition is even more challenging than in the case of superposed slits. Possible interpretation could be that at every instance of time inside the process $\ket{(A+B)/\sqrt{2}}_{process}$ the particle ``sees" the space in the same way it ``sees" it when it passes the slits in the analogous superposition $\ket{(A+B)/\sqrt{2}}_{slit}$. This is the cost of being able to order the ICO processes: the processes become delocalized and it is difficult to understand what each of them does, but this does not undermine the fact that ICO processes may be ordered. In fact, such a tradeoff has Bohr-like complementary feeling to it. 

\section{Conclusions}

We have presented a way to order ICO processes. Therefore we believe that ICO processes cannot be considered an argument in favour of fundamental unorderability or acausality of quantum mechanics and of nature and this question has to wait for frame-of-reference independent arguments. In fact, very rarely, when speaking about ICO, particularity of frame of reference is discussed at all (see \cite{7} for an example of explicit statement of the frame of reference of ICO processes and \cite{10} for the introduction of Quantum Equivalence Principle in the context of ICO). It seems that vagueness of our everyday words when we interpret quantum superposition extends to vagueness when we interpret more intricate consequences of quantum superposition, creating paradoxical situations. The more we try to interpret quantum mechanics, the more is evident the incompatibility of logical structures at the root of our current language with the logic of quantum mechanics. Perhaps quantum mechanics has to enter the area of interest of linguists in order to overcome this difficulty.



\end{document}